\documentclass[runningheads]{llncs}

\usepackage{eccv}

\usepackage{eccvabbrv}

\usepackage{graphicx}
\usepackage{booktabs}
\usepackage{algorithm}
\usepackage{algpseudocode}

\usepackage[accsupp]{axessibility}  

\usepackage{xcolor}

\usepackage[pagebackref,breaklinks,colorlinks,citecolor=eccvblue]{hyperref}

\usepackage{orcidlink}

\newcommand{\gsrc}{\mathcal{G}^\text{src}}
\newcommand{\gedt}{\mathcal{G}^\text{edit}}
\newcommand{\gft}{\mathcal{G}_\text{ft}}

\newcommand{\y}{y}
\newcommand{\Isrc}{\mathcal{I}^\text{src}}
\newcommand{\Iedt}{\mathcal{I}^\text{edit}}
\newcommand{\Vs}{\mathcal{V}}
\newcommand{\G}{\mathcal{G}}
\newcommand{\zs}{z^\text{src}}
\newcommand{\zo}{z^\text{ori}}
\newcommand{\ze}{z^\text{edit}}
\newcommand{\zc}{z^\text{con}}
\newcommand{\zb}{z^\text{bld}}

\newcommand{\dec}{\mathcal{D}}
\newcommand{\m}{\mathbf{M}}
\newcommand{\mc}{\mathbf{M}^\text{con}}
\newcommand{\mm}{\mathbb{M}}
\newcommand{\eps}{\varepsilon}

\definecolor{tabhighlight}{gray}{0.9}
\begin{document}

\title{View-Consistent 3D Editing with Gaussian Splatting} 

\titlerunning{View-Consistent 3D Editing with Gaussian Splatting}

\author{Yuxuan Wang$^{1}$, Xuanyu Yi$^{1,2}$, Zike Wu$^1$, Na Zhao$^3$, Long Chen$^5$, and Hanwang Zhang$^{1,4}$}

\institute{$^1$Nanyang Technological University \quad $^2$Institute for Infocomm Research, A*STAR \\ $^3$Singapore University of Technology and Design  \quad $^4$Skywork AI \\ $^5$Hong Kong University of Science and Technology}

\authorrunning{Y. Wang et al.}

\maketitle

\begin{figure}[h]
\centering
\includegraphics[width=0.99\linewidth]{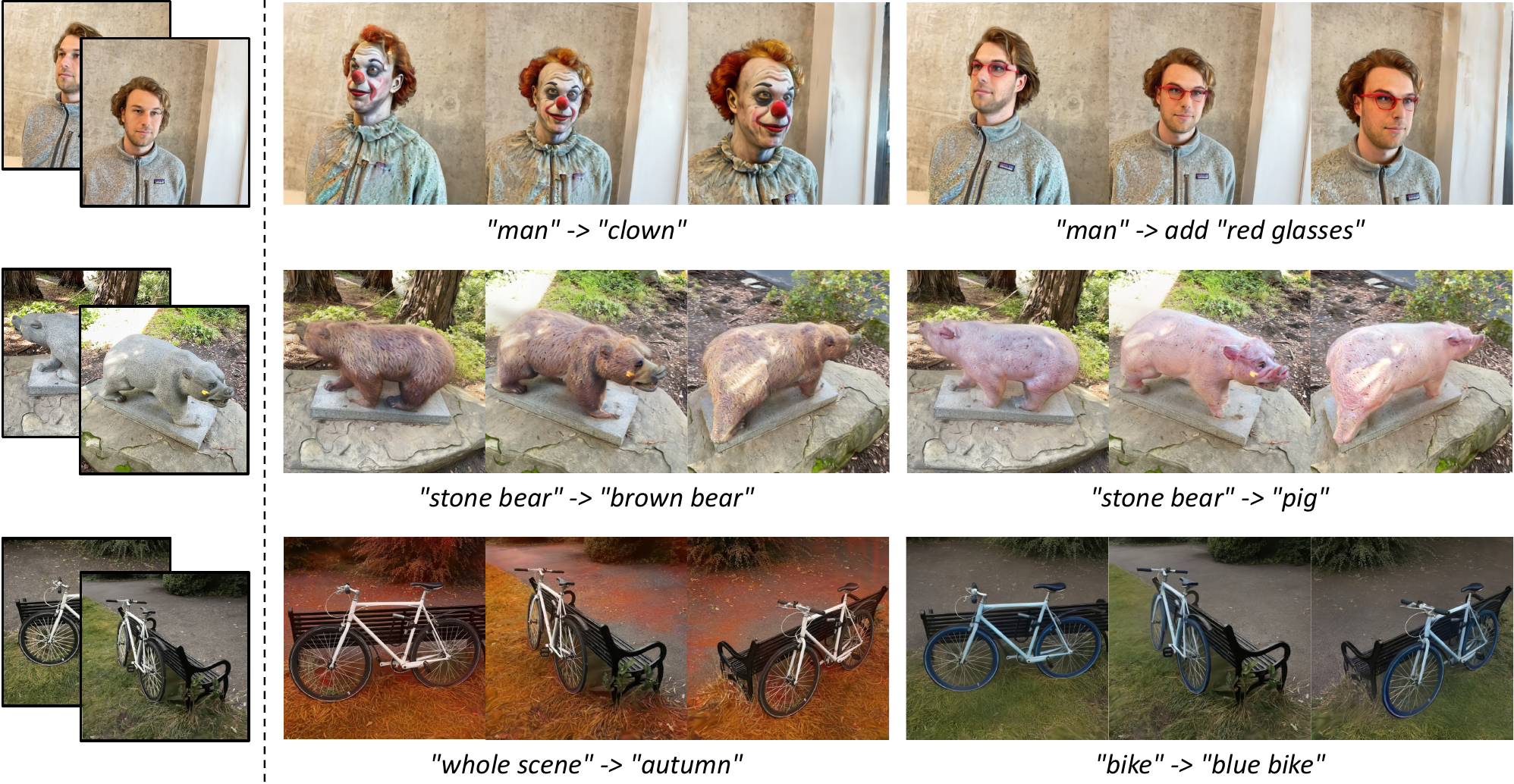}
\caption{Capability highlight of our method: \textsc{VcEdit}. Given a source 3D Gaussian Splatting and a user-specified text prompt, our \textsc{VcEdit} enables versatile scene and object editing. By ensuring multi-view consistent image guidance, \textsc{VcEdit} alleviates artifacts and excels in high-quality editing.}
\label{fig:fig1}
\end{figure}

\begin{abstract}

The advent of 3D Gaussian Splatting (3DGS) has revolutionized 3D editing, offering efficient, high-fidelity rendering and enabling precise local manipulations. 
Currently, diffusion-based 2D editing models are harnessed to modify multi-view rendered images, which then guide the editing of 3DGS models.
However, this approach faces a critical issue of multi-view inconsistency, where the guidance images exhibit significant discrepancies across views, leading to mode collapse and visual artifacts of 3DGS. 
To this end, we introduce View-consistent Editing (\textsc{VcEdit}), a novel framework that seamlessly incorporates 3DGS into image editing processes, ensuring multi-view consistency in edited guidance images and effectively mitigating mode collapse issues. 
\textsc{VcEdit} employs two innovative consistency modules: the Cross-attention Consistency Module and the Editing Consistency Module, both designed to reduce inconsistencies in edited images. 
By incorporating these consistency modules into an iterative pattern, \textsc{VcEdit} proficiently resolves the issue of multi-view inconsistency, facilitating high-quality 3DGS editing across a diverse range of scenes. The project page and codes are shown in
\href{http://vcedit.github.io}{http://vcedit.github.io}.

  \keywords{3D Editing \and 3D Gaussian Splating \and Multi-view Consistency \and Diffusion Model}
\end{abstract}

\section{Introduction}
\label{sec:intro}

We consider the problem of text-driven 3D model editing: given a source 3D model and user-specified text instructions, the task is to modify the 3D model according to the instructions, as depicted in Fig.~\ref{fig:fig1}, ensuring both editing fidelity and preservation of essential source content~\cite{song2023efficient,chen2023plasticine3d,cheng2023progressive3d,chen2023shap,karim2023free}.
This problem holds paramount importance across a variety of industrial applications, such as real-time outfit changes for 3D digital humans and immersive AR/VR interactive environments\cite{yi2023invariant,cao2023dreamavatar,yuen2011augmented}. 
Recently, groundbreaking 3D Gaussian Splatting (3DGS)~\cite{kerbl20233d} has emerged as a promising ``silver bullet'' for 3D editing, notable for its efficient, high-fidelity rendering and explicit representation (3D anisotropic balls known as \textit{Gaussians}) suitable for local manipulation.  

\begin{figure}[t]
\centering
\includegraphics[width=0.99\linewidth]{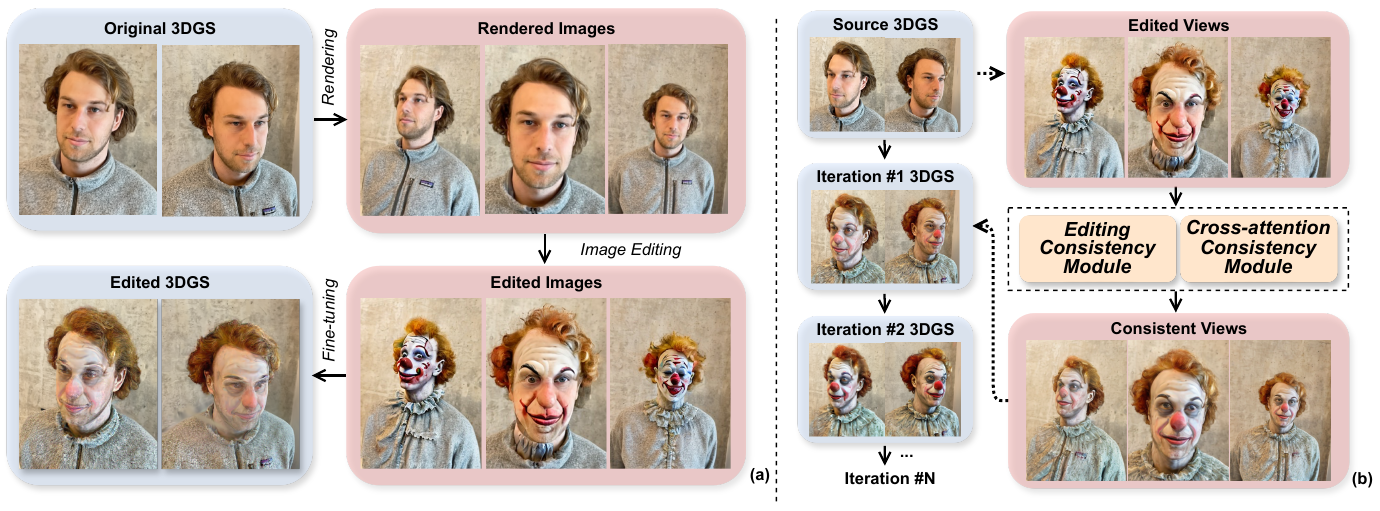}
\caption{\textbf{(a)}: Current image-guided 3DGS editing pipeline and its \textbf{multi-view inconsistency issue}: The rendered views of a man are separately edited to different styles of clowns, leading to the mode collapse issue of learned 3DGS.
\textbf{(b)}: The iterative pattern and our consistency modules deployed in each iteration. The 3DGS is progressively guided to a coherent style that aligns with the ``clown'' through the iterative pattern.}
\label{fig:inconsistency}
\end{figure}

In the context of editing 3DGS models, thanks to recent progress in large-scale pre-trained 2D diffusion models, existing methods~\cite{haque2023instructnerf2nerf,dihlmann2024signerf,dong2024vicanerf,chen2023gaussianeditor,fang2023gaussianeditor} leverage off-the-shelf 2D editing models~\cite{gal2022image,gan2023instructcv,parmar2023zeroshot,xu2023inversionfree,feng2024item, bai2023integrating} to guide optimization of the 3DGS model.
As shown in Fig.~\ref{fig:inconsistency}(a), this pattern renders source 3DGS into multi-view 2D images, manipulates them via 2D editing models using text prompts, and then employs these adjusted images to fine-tune the original 3DGS. 
Beyond achieving plausible editing outcomes, this image-based pattern also facilitates user-friendly interaction, enabling users to pre-select their preferred edited images and personalize the editing workflow.

However, such image-guided 3DGS editing has a notorious \textbf{multi-view inconsistency} that cannot be ignored.
Fig.~\ref{fig:inconsistency}(a) vividly illustrates that images edited separately using a state-of-the-art 2D editing model~\cite{xu2023inversionfree} manifest pronounced inconsistencies across views --- the views of a man are edited to different styles of clowns.
Utilizing these significantly varied edited images as guidance, the 3DGS model will struggle with the issue that few training images are coherent, whereas the majority display conflicting information.
Unfortunately, the explicitness and inherent densification process of 3DGS make it especially vulnerable to multi-view inconsistency, 
which complicates 3DGS in densifying under-reconstruction regions or pruning over-reconstruction regions~\cite{tang2023dreamgaussian}.
Consequently, training with the multi-view inconsistent guidance can lead to the mode collapse of 3DGS, characterized by the ambiguity between the source and target, as well as the flickering artifacts revealed in Fig.~\ref{fig:inconsistency}(a).

Thus, the crux lies in addressing the multi-view inconsistency of the image guidance.
We conjecture such a problem stems from the lack of 3D awareness in 2D editing models; that is, they inherently process each view in isolation.
Therefore, we introduce the \textbf{View-consistent Editing} (\textsc{VcEdit}), a high-quality image-guided 3DGS editing framework. This framework seamlessly incorporates 3DGS into image editing processes to achieve multi-view consistent guidance, thus effectively addressing the issue of 3DGS mode collapse.
As illustrated in Fig.~\ref{fig:inconsistency}(b), \textsc{VcEdit} employs specially-designed multi-view consistency modules within an iterative pattern.

Primarily, we design two effective consistency modules using the explicit nature and fast rendering capability of 3DGS:
(1) The \textbf{Cross-attention Consistency Module (CCM)} that consolidates the multi-view cross-attention maps in the diffusion-based image editing model, thus harmonizing the model's attentive 3D region across all views. More concretely,
this process inverse-renders the original cross-attention maps from all views onto each Gaussian within the source 3DGS, thereby creating an averaged 3D map.
Subsequently, this 3D map is rendered back to 2D, serving as the consolidated cross-attention maps to replace the originals for more coherent edits. 
(2) The \textbf{Editing Consistency Module (ECM)} that directly calibrates the multi-view inconsistent editing outputs: We fine-tune a source-cloned 3DGS with the editing outputs and then render the 3DGS back to images. Taking advantage of the rapid rendering speed of the 3DGS, this mechanism efficiently decreases incoherent content in each edited image.


To further mitigate the multi-view inconsistency issue, we extend our \textsc{VcEdit} to an \textbf{iterative pattern}: editing rendered images $\rightarrow$ updating the 3DGS $\rightarrow$  repeating.
In situations where the image editing model yields overly inconsistent initial edits, it allows for the correction of initially inconsistent views in later iterations \cite{haque2023instructnerf2nerf}, which continuously refines the 3DGS and fosters a reciprocal cycle.
 Fig.~\ref{fig:inconsistency}(b) illustrates that the 3DGS of a ``man'' is iteratively guided to a consistent style that aligns with the desired ``clown'' target.
To meet the demand for rapid iteration, \textsc{VcEdit} integrates InfEdit \cite{xu2023inversionfree}, a high-quality, fast image editing model that bypasses the lengthy DDIM-inversion phase. Depending on the complexity of the editing scene and user instructions, the processing time in our \textsc{VcEdit} for each sample ranged from 10 to 20 minutes.

As illustrated in Fig.~\ref{fig:fig1}, by incorporating consistency modules and the iterative pattern, \textsc{VcEdit} significantly enhances the multi-view consistency in guidance images, leading to superior editing quality. 
We conducted comprehensive evaluations of \textsc{VcEdit} in various real-world scenes.
Both qualitative and quantitative experiments clearly indicate that \textsc{VcEdit} can address the mode collapse of 3DGS, continuously outperforming other state-of-the-art methods. 
Our contributions can be summarized in three aspects:
\begin{itemize}
\setlength{\itemsep}{2pt}
\setlength{\parsep}{0pt}
\setlength{\parskip}{0pt}

\item We propose effective multi-view \textbf{consistency modules} that harmonize the inconsistent multi-view guidance images.
\item Based on our designed consistency module, we construct an effective \textbf{iterative pattern} that offers high-quality 3DGS editing.
\item We pioneer a paradigm that integrates 3DGS characteristics with multi-view consistency design, inspiring further efforts and explorations.
\end{itemize}

\section{Related Works} 


\noindent
\textbf{Text-guided Image Editing.}
Large-scale pretrained Text-to-image (T2I) diffusion models~\cite{rombach2022high,wu2023fast} have been widely adopted in generative image editing, guided by specific text prompts. Textual Inversion ~\cite{gal2022image} optimizes special prompt token(s) in text embedding space to represent the specified concept, whereas Prompt-to-Prompt leverages cross-attention information to maintain integrity in unmodified regions. Pix2pix-zero~\cite{parmar2023zeroshot} utilizes embedding vector mechanisms to establish controllable editing. InstructPix2Pix~\cite{gan2023instructcv} merges the capacities of GPT and Stable Diffusion~\cite{rombach2022high} to generate a multi-modal dataset and fine-tunes Stable Diffusion to achieve instructed diverse editing. Our \textsc{VcEdit} integrates InfEdit~\cite{xu2023inversionfree}, which implies a virtual inversion strategy and attention control mechanisms for a controllable and faithful editing. Nevertheless, all the above works mainly focus on editing a single image, and obtaining high-quality paired data with multi-view consistency from the 2D diffusion model is extremely challenging, resulting in the application of 3D-aware editing impractical.

\noindent
\textbf{3D Field Editing.} In the last few years, NeRF-based approaches~\cite{mikaeili2023sked,richardson2023texture,kamata2023instruct,yu2023edit,zhou2023repaint,park2023ed,khalid2023latenteditor,bao2023sine,karim2023free,yi2024diffusion} have overtaken traditional point cloud and triangle mesh approaches~\cite{zhang2023point,wang2023meshguided,zhang2024dragtex,park2023ed, bao2023sine} for 3D field editing. DreamEditor~\cite{zhuang2023dreameditor} updates textual attention areas with score distillation sampling (SDS) guidance~\cite{poole2022dreamfusion}. DDS~\cite{hertz2023delta} and PDS~\cite{koo2023posterior} introduce an advance SDS loss with cleaner optimization direction for better instruction alignment. FocalDreamer~\cite{li2023focaldreamer} generates new geometries in specified empty spaces from the text input. Instruct-NeRF2NeRF~\cite{haque2023instructnerf2nerf} iteratively refines dataset images using the InstructPix2Pix diffusion model and NeRF, a pattern widely adopted in subsequent research~\cite{dong2024vicanerf,li2024instructpix2nerf}. However, the inherent computational demands and the implicit nature of NeRF usually require time-consuming fine-tuning for edit operations and lack of precise controllability, which cannot support practical interactive editing use cases. More recent works~\cite{fang2023gaussianeditor,chen2023gaussianeditor} integrate Gaussian splatting into 3D field editing and surpasses previous 3D editing methods in terms of effectiveness, speed, and controllability. Our \textsc{VcEdit} builds upon the above image-based 3DGS editing approaches and further harmonizes the inconsistent multi-view image guidance they overlook, which enables more high-fidelity editing quality without mode collapse.

\section{Preliminary}



\subsection{3D Gaussian Splatting}
3D Gaussian splatting (3DGS) has emerged as a prominent efficient explicit 3D representation technique, employing anisotropic 3D Gaussians for intricate modeling~\cite{kerbl20233d}. Each Gaussian, represented as $G$, is characterized by its mean $\mu \in \mathbb{R}^3$, covariance matrix $\Sigma$, associated color $c \in \mathbb{R}^3$, and opacity $\alpha \in \mathbb{R}$. For better optimization, the covariance matrix $\Sigma$ is decomposed into a scaling matrix $S \in \mathbb{R}^3$ and a rotation matrix $R \in \mathbb{R}^{3\times3}$, with $\Sigma$ defined as $\Sigma = R S S^T R^T$.
A Gaussian centered at $\mu$ is expressed as $G(x) = \exp\left(-\frac{1}{2} x^T \Sigma^{-1} x\right)$, where $x$ denotes the displacement from $\mu$ to a point in space. 
In the splatting rendering process, 
we render the color $c$ for a pixel by blending all sampled 3D points along the ray going through this pixel:
\begin{equation}
c = \sum_{i} c_i \alpha_i G(x_i) \prod_{j=1}^{i-1} \left(1 - \alpha_j G(x_j)\right).
\end{equation}

An efficient tile-based rasterizer~\cite{kerbl20233d} allows for rapid forward and backward passes, supporting high-quality real-time rendering. The optimization of 3D Gaussian properties is conducted in tandem with an adaptive \textit{density control mechanism}, involving densifying and pruning operations, where Gaussians are added and occasionally removed. By incorporating all the Gaussians $G$, the 3DGS model, denoted as $\mathcal{G}$, represents complete 3D scenes and objects.

\subsection{Image-guided 3DGS Editing}
\label{3dgs editing}

Given a source 3D Gaussian Splatting (3DGS) model $\gsrc$ and a user-specified target prompt $\y$, the goal of 3DGS editing is to transform $\gsrc$ into an edited version $\gedt$ that aligns with $\y$. 
In image-guided 3DGS editing, 
$\gsrc$ is first rendered from multiple views $\Vs = \{v\}$, generating a collection of source images $\Isrc$. Subsequently, a 2D editing model transforms $\Isrc$ into edited images $\Iedt$ according to $\y$. 
Finally, these edited images serve as training guidance to refine $\gsrc$ into the edited version $\gedt$. 
Specifically, an editing loss $L_\text{Edit}$ is calculated for each view between the real-time rendering and the edited images. 
The final edited 3DGS model is obtained by minimizing the editing loss across all views $\mathcal{V}$:
\begin{equation}
\gedt = \: \underset{\G}{\mathrm{argmin}} \: \sum_{v \in \Vs}{ \mathcal{L_\text{Edit}}(\mathcal{R}(\G, v),\: \Iedt) },
\end{equation}
where $\mathcal{R}$ represents the rendering function that projects 3DGS to image given a specific view $v$. However, as introduced in Section~\ref{sec:intro}, the edited images $\Iedt$ produced by 2D editing models exhibit severe multi-view inconsistency issue if no further control is conducted. This inconsistency in 2D guidance images will lead to mode collapse and low-quality editing in the edited 3DGS model.
\section{Our Method}

In this section, we introduce the View-Consistent Editing~(\textsc{VcEdit}), a novel approach for 3DGS editing.
First, we outline the general pipeline of \textsc{VcEdit} in Sec.~\ref{sec:imgedit}. 
Subsequently, we introduce two innovative Consistency Modules designed to address the multi-view inconsistency issue.
Particularly, in Sec.~\ref{sec:acc}, we introduce the Cross-attention Consistency Module, which aims to reduce multi-view editing variance by improving the attention consistency. 
In Sec.~\ref{sec:pcc}, we present the Editing Consistency Module, which directly calibrates multi-view editing predictions to ensure consistency across all predictions. 
Finally, in Sec.~\ref{sec:iter}, we extend our framework to an iterative pattern, which facilitates the iterative improvement of consistency between $\Iedt$ and $\gedt$.



\subsection{View-consistent Editing Pipeline}
\label{sec:imgedit}

\begin{figure}[t]
\centering
\includegraphics[width=0.99\linewidth]{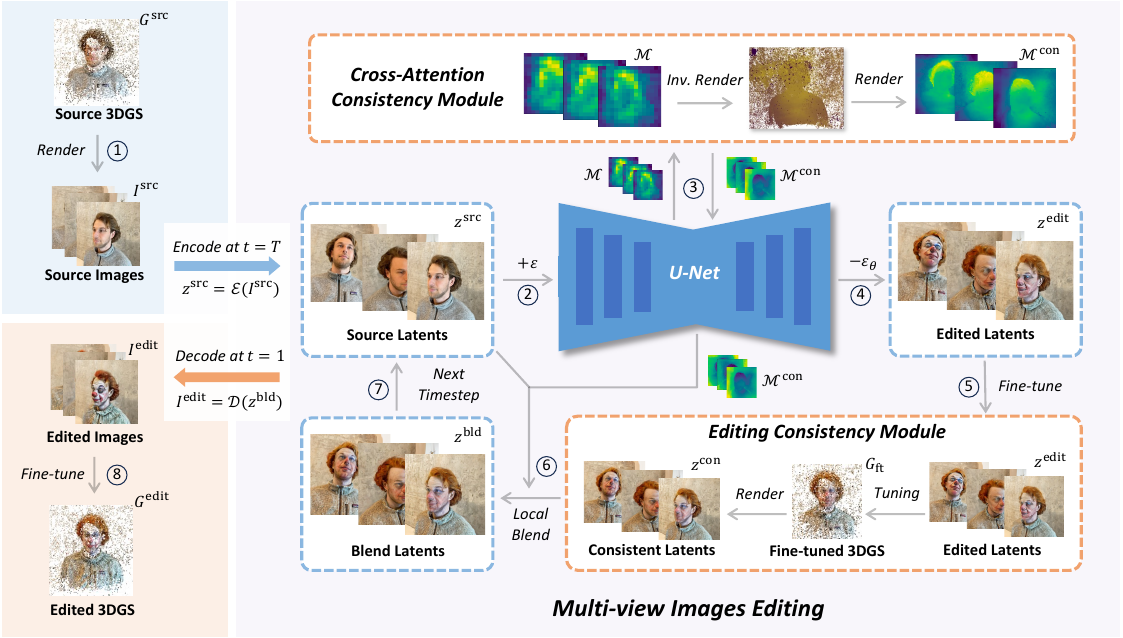}
\caption{The pipeline of our \textsc{VcEdit}: \textsc{VcEdit} employs an image-guided editing pipeline. In the image editing stage, the Cross-attention Consistency Module and Editing Consistency Module are employed to ensure the multi-view consistency of edited images. We provide a detailed overview in Sec.~\ref{sec:imgedit}.}

\label{fig:method}
\end{figure}

Fig.~\ref{fig:method} illustrates the overview of our framework, \textsc{VcEdit}. To
perform image-guided 3DGS editing, we utilize the pre-trained 2D diffusion model \cite{rombach2022high,wu2023fast} to generate a series of edited images given a set of multi-view rendered images.

Inspired by~\cite{xu2023inversionfree}, we adopt an annealed timestep schedule from the largest timestep $T$ to $1$, where we evenly sample $N$ timesteps from the schedule, denoted as $t \in \{t_i\}_{i=1}^N$.
Before the start of the first timestep $t_1$, the rendered images, denoted as $\Isrc$, are encoded into multi-view source latents $\zs$.

For each sampled timestep, the source latents $\zs$ go through an editing process.
Specifically, at timestep $t$, we first adopt the forward diffusion process to perturb the given rendered images with random Gaussian noise $\eps$ as follows:
\begin{equation}
    z_t = \sqrt{\alpha_t}\zs + \sqrt{1 - \alpha_t} \eps, \: \eps \sim \mathcal{N}(\mathbf{0}, \mathbf{I}),
\end{equation}
where the $\alpha_t$ is a time-variant hyper-parameter, $\zs$ is the encoded latent images. 
Then we feed the noisy latents $z_t$ into a pre-trained diffusion model $\eps_{\theta}$ to generate denoised images as edited latents based on the user's target prompt $y$:
\begin{equation}
    \ze = (z_t - \sqrt{1 - \alpha_t} \cdot \eps_\theta(z_t, \y)) / \sqrt{\alpha_t}.
\end{equation}




However, we observe that the edited images generated by simply adopting pre-trained 2D diffusion models lacks of multi-view consistency, which often leads to mode collapse in the learned 3DGS~\cite{liu2024one}. 
Therefore, we propose the Cross-attention Consistency Module in Sec.~\ref{sec:acc}, which consolidates multi-view cross-attention maps within each layer of the diffusion model's U-Net architecture, improving the view-consistency in $\ze$.

Upon obtaining the edited latents, we aim to further improve the multi-view consistency of the edited guidance images.
To this end, we design the Editing Consistency Module, where we adopt a fine-tuning and rendering process using 3DGS to calibrate $\ze$, producing a set of more consistent latents, denoted as $\zc$. 
Finally, a local blending process is conducted to preserve source information, generating the blended latents, denoted as $\zb$, which serve as the output of the current timestep $t$. 
Details are demonstrated in Sec.~\ref{sec:pcc}.

Upon completion of one timestep, its output, $\zb$, continually serves as the input of the succeeding timestep. This editing process is repeated as $t$ progresses from $t_1$ to $t_N$. Finally, the last timestep's output are decoded back to multi-view edited images, denoted as $\Iedt$, which serves as the view-consistent guidance images to fine-tune the $\gsrc$, where the training objective includes a MAE loss and a VGG-based LPIPS loss~\cite{zhang2018unreasonable,simonyan2014very}:
\begin{equation}
    \mathcal{L} = \lambda_1\mathcal{L}_{\text{MAE}}(\mathcal{R}(\G, v), \Iedt) + \lambda_2\mathcal{L}_{\text{LPIPS}}(\mathcal{R}(\G, v), \Iedt)
\end{equation}


In summary, \textsc{VcEdit} is integrated with two effective consistency modules. In Sec.~\ref{sec:ablation_consist}, our experiments thoroughly demonstrate that it can produce consistent image editing results which are directly used as guidance.

\subsection{Cross-attention Consistency Module}
\label{sec:acc}

As discussed in Sec.~\ref{sec:intro}, existing image-guided 3DGS editing approaches~\cite{haque2023instructnerf2nerf,fang2023gaussianeditor,chen2023gaussianeditor} frequently encounter multi-view inconsistencies in the generated guidance images.
We suggest that this inconsistency arises from the lack of cross-view information exchange during the editing process.

Therefore, we develop the \textbf{Cross-attention Consistency Module (CCM)} that consolidates the cross-attention maps from all views during the forward pass of the U-Net, where these maps indicate the pixel-to-text correlations between images and user prompts.
In this way, we successfully enable cross-view information exchange within the backbone network in a seamless, plug-and-play manner, thereby facilitating multi-view consistency in the generation of edited images.



As illustrated in the \textit{top} of Fig.~\ref{fig:method}, we first employ inverse-rendering~\cite{kerbl20233d, chen2023gaussianeditor} to map the 2D cross-attention maps $\m$ back onto the 3D Gaussians within $\gsrc$, generating a 3D map $\mm$.
Given the $j$-th Gaussian in $\gsrc$ and the pixel $\textbf{p}$ in the $v$-th 2D map $\m_v$, the $o_j(\mathbf{p})$, $T_{j, v}(\mathbf{p})$, $\m_{v, \mathbf{p}}$ denote the opacity, transmittance matrix, and cross-attention weight, respectively.
Specifically, for the $j$-th Gaussian in $\gsrc$, its value in the 3D map $\mm_j$ is computed as: 
\begin{equation}
    \mm_j = \frac{1}{C_j}\sum_{v \in V} \sum_{\textbf{p} \in \m_v} o_j(\mathbf{p}) \cdot T_{j, v}(\mathbf{p}) \cdot \m_{v, \mathbf{p}}
\end{equation}
The $C_j$ is the count indicating the total number of values that assigned to $j$ by $\m_{v, \mathbf{p}}$ across all $\mathbf{p}$ and $v$. 
In the created 3D map $\mm$, each Gaussian receives an attention weight towards each word in the prompt.
Subsequently, we render the $\mm$ back to 2D space to obtain a set of consistent cross-attention maps, denoted as $\mc$.

By replacing the vanilla cross-attention maps $\m$ with the consistent attention maps $\mc$ 
, our CCM conducts an 3D average pooling to unify model's attentive region across different views. 
In this way, the U-Net model ensures uniform attention is given to the same 3D region across different views during the editing process, facilitating the production of more consistent edited latents.
Our experiments, detailed in Sec.~\ref{sec:ablation_consist}, validate that this cross-attention control mechanism significantly bolsters view consistency.







\subsection{Editing Consistency Module}
\label{sec:pcc}


Incorporating the CCM described in Sec.~\ref{sec:acc} allows the diffusion models to produce $z^{\text{edit}}$ with enhanced consistency. Yet, it is still inadequate for updating the 3DGS directly, since 
3DGS is based on spatial physical modeling that requires coherent representation across views.
Therefore, we introduce \textbf{Editing Consistency Module (ECM)} which distills the image guidance into an intermediate model incorporating physical modeling and consistency constraints. 
This intermediate model is then utilized to generate more coherent multi-view edited guidance images.

We illustrate the proposed module in the \textit{bottom-right} of Fig.~\ref{fig:method}. Initially, we decode the multi-view edited latents $\ze$ into images, denoted as $\dec(\ze)$, which are subsequently used as guidance to fine-tuned a copy of the original 3DGS model $\gsrc$, which is denoted as $\gft$:
\begin{equation}
\gft = \: \underset{\G}{\mathrm{argmin}} \: \sum_{v \in \Vs}{ \mathcal{L}_{\text{edit}}(\mathcal{R}(\G, v),\: \dec(\ze_{v})) }
\end{equation}

Following the rapid fine-tuning process, we render the fine-tuned model $\mathcal{G}_{\text{ft}}$ from each viewpoint $v \in \mathcal{V}$, yielding a collection of rendered images that exhibit significantly enhanced coherence. Finally, these images are encoded to the consistent latents, denoted as $\zc$.

Afterwards, we follow \cite{xu2023inversionfree} and employ a Local Blend module to preserve the unedited details in the source image:
Given the cross-attention maps $\mc$, the source latents $\zs$, and the consistent latents $\zc$, the blending is conducted by:
\begin{equation}
\zb = \mc * \zc  + (1 - \mc) * \zs
\end{equation}


Our empirical findings in \textit{appendix} suggest that 3DGS is capable of rectifying inconsistencies when the discrepancies between views are minor. Our image editing process, unfolding across various diffusion timesteps following InfEdit~\cite{xu2023inversionfree}, guarantees that $\ze$ undergoes only slight changes from $\zs$, maintaining a low level of inconsistencies at each step.
Consequently, through continuously calibration at each timestep, the ECM effectively prevents the inconsistencies from accumulating.
In our ablation study (Sec.~\ref{sec:ablation_consist}), we demonstrate the significant capability of this mechanism to address multi-view inconsistencies.



\subsection{Iterative Pattern Extension}
\label{sec:iter}
To further alleviate the multi-view inconsistency issue, we evolve our \textsc{VcEdit} into an iterative pattern~\cite{yi2022identifying}.
Based on the image-guided editing pipeline detailed in Sec.~\ref{3dgs editing}, we define each sequence of transitions ---from $\gsrc$ to $\Isrc$, from $\Isrc$ to $\Iedt$, and from $\Iedt$ to $\gedt$---as an iteration.
Upon completion of one iteration, the $\gedt$ is forwarded to the subsequent iteration and serves as the new $\gsrc$.

In this pattern, each iteration incrementally aligns the 3DGS more closely with the user's prompt, facilitating smoother image editing.
This fosters a mutual cycle, where $\Iedt$ and $\gedt$ successively refine each other, resulting in ongoing enhancements of editing quality~\cite{haque2023instructnerf2nerf}.
This leads to more consistent and higher-quality guidance images, which in turn refines the 3DGS for improved $\gedt$.
Taking advantage of our two effective Consistency Modules proposed in Sec.~\ref{sec:acc} and Sec.~\ref{sec:pcc}, only a few iterations are needed to achieve satisfying editing results.
By integrating these strategies, our \textsc{VcEdit} framework effectively ensures multi-view consistency in guidance images and excellence in 3DGS editing.

\section{Experiments}



\subsection{Implementation Details}
\label{sec:imple}
We implement our experiments using PyTorch based on the official codebase of 3D Gaussian splatting \cite{kerbl20233d}.
All of our experiments are conducted on a single NVIDIA RTX A6000 GPU.
We used Adam optimizer~\cite{kingma2014adam} with a learning rate of 0.001 in both fine-tuning the source 3DGS and our Editing Consistency Module.  
For each 3DGS editing task, we conduct one to three editing iterations where the source 3DGS is optimized for 400 steps.


\subsection{Baselines and Evaluation Protocol}
\label{sec:expset}
We compare our proposed \textsc{VcEdit} method with two baselines: (1) \textbf{Delta Distillation Sampling} (DDS)~\cite{hertz2023delta}, which utilizes an implicit score function loss derived from a diffusion model as guidance, and (2) the state-of-the-art \textbf{GSEditor}~\cite{chen2023gaussianeditor}, which employs image-guided editing and iteratively updates a training set to steer 3DGS modifications. To ensure a fair comparison, all methods incorporate the same local 3DGS update process pioneered by GSEditor.
Our quantitative comparisons assess the similarity of appearance between the edited 3DGS and the user's prompt with the CLIP-similarity metrics~\cite{gal2022stylegan}. Furthermore, a user study was conducted to evaluate the effectiveness of editing methods from a human perspective. Please refers to \textit{Appendix} for the detailed configuration of the user study. 

\begin{figure}[t]
\centering
\includegraphics[width=0.99\linewidth]{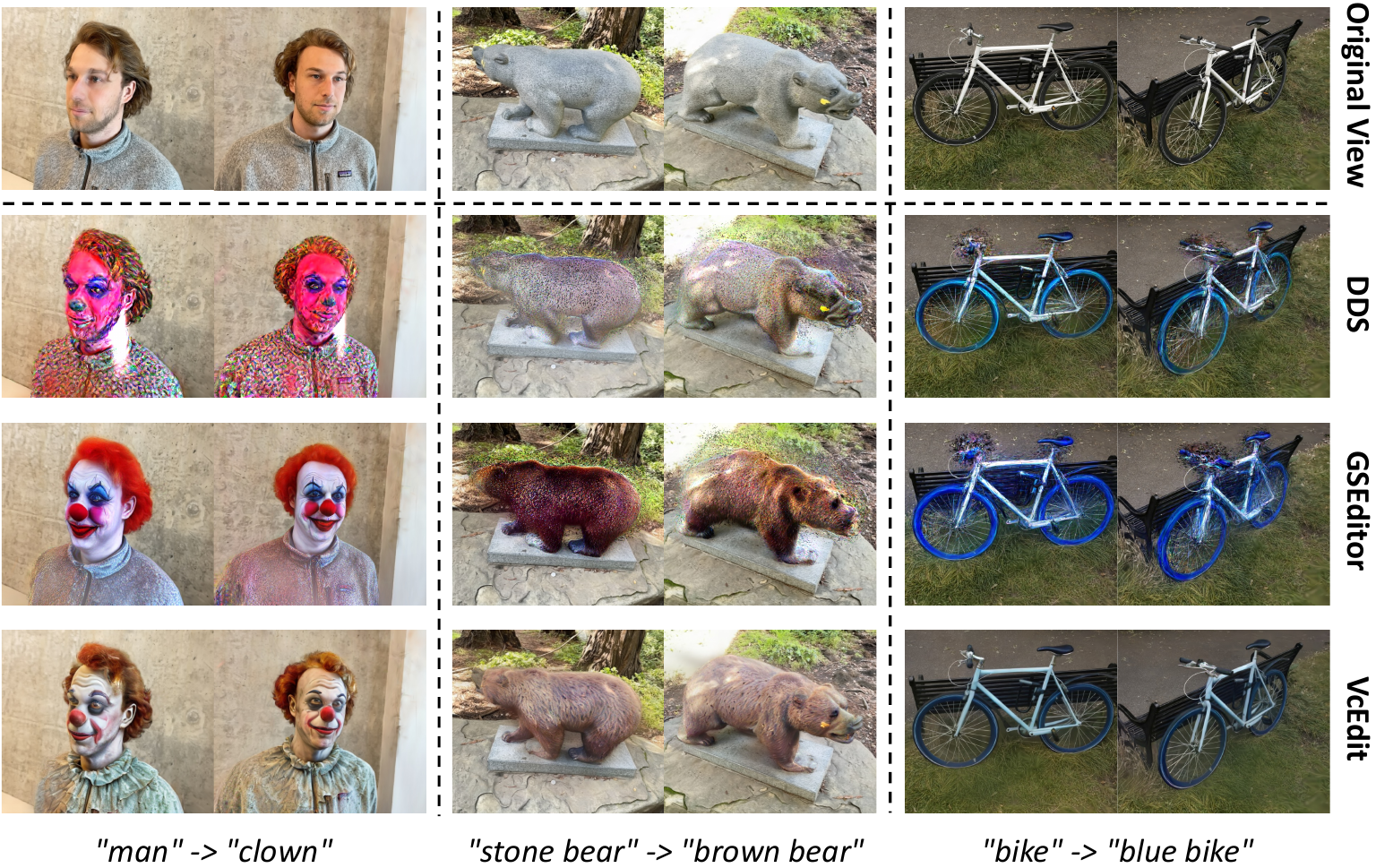}
\caption{Qualitative comparison with the DDS~\cite{hertz2023delta} and the GSEditor~\cite{chen2023gaussianeditor}: 
The \textit{topmost} row demonstrate the original views, while the \textit{bottom} rows show the rendering view of edited 3DGS.
\textsc{VcEdit} excels by effectively addressing the multi-view inconsistency, resulting in superior editing quality. In contrast, other methods encounter challenges with mode collapse and exhibit flickering artifacts.}
\label{fig:sota}
\end{figure}

\begin{figure}[t]
\centering
\includegraphics[width=0.99\linewidth]{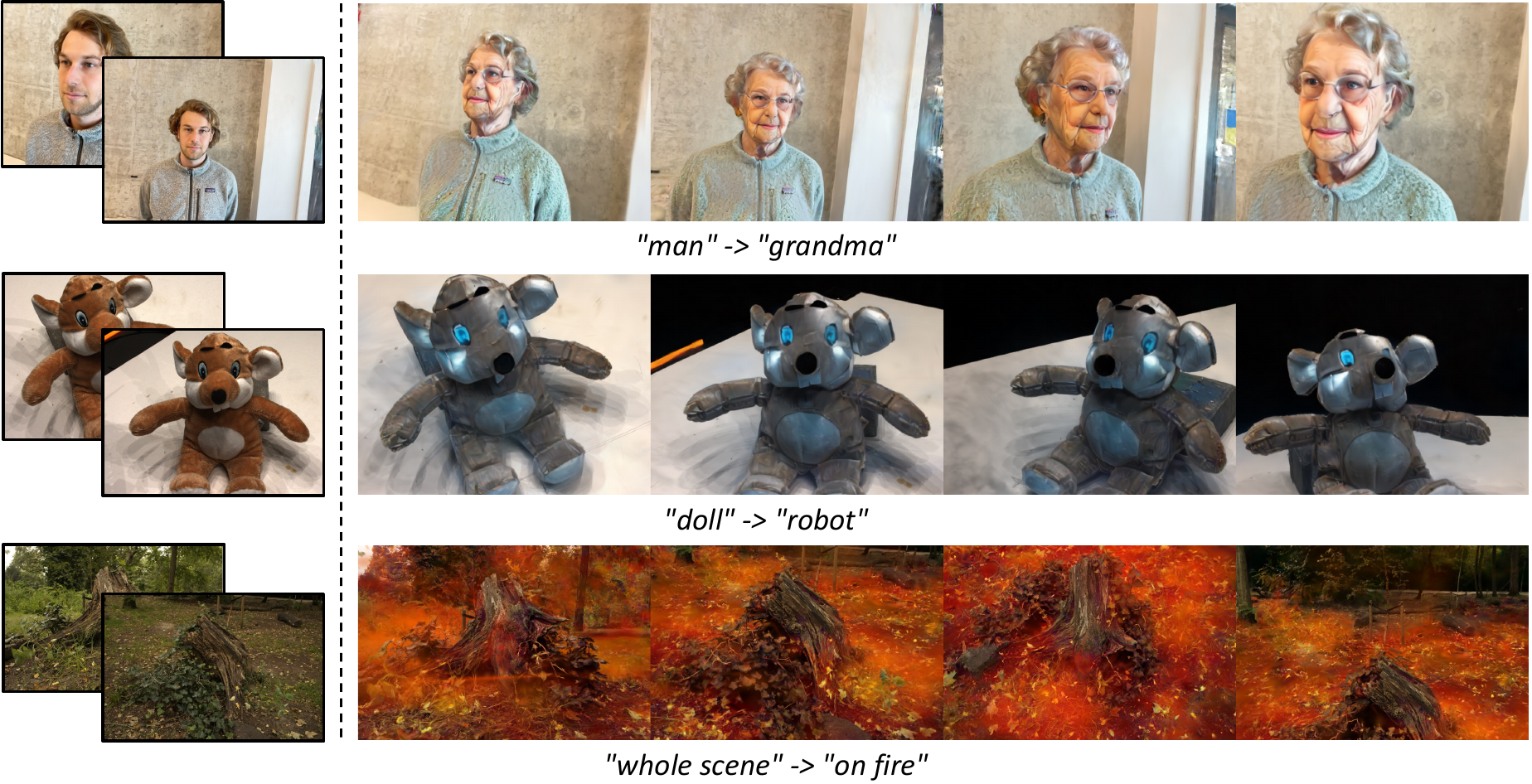}
\caption{Extensive Results of \textsc{VcEdit}: Our method is capable of various editing tasks, including face, object, and large-scale scene editing. The leftmost column demonstrates the original view, while the right four columns show the rendered view of edited 3DGS.}
\label{fig:extensive}
\end{figure}

\subsection{Qualitative Comparisons}
\label{sec:quali}
We meticulously select three challenging scenarios ranging from human face editing to outdoor scene editing. In the first column of Fig.~\ref{fig:sota}, the precision-demanding task of human face editing highlights the critical importance of accurate manipulation due to the sensitivity of facial features to imperfections. Subsequent columns feature objects placed within outdoor scenes with panoramic rendering views, posing a challenge for methods to maintain view-consistent guidance during editing.

Fig.~\ref{fig:sota} shows the editing results achieved by our \textsc{VcEdit} along with those of the baseline methods.
(1) We observe a notable editing failure in all the DDS results, where the 3DGS presents severe noisy coloration and shape distortion.
(2) When compared to DDS, the output from GSEditor shows better alignment with the target prompt. 
However, the resulting 3DGS exhibits noticeable artifacts, including noisy coloration and blurred edges.
(3) Unlike DDS and GSEditor, our \textsc{VcEdit} consistently delivers clean and intricately detailed edits in all scenes.
Our results confirm that \textsc{VcEdit} exhibits superior editing performance compared to other baselines. 

\begin{table}[h]
\centering
\caption{Quantitative Comparison: Our \textsc{VcEdit} performs in both user study evaluations and CLIP T2I Directional Similarity~\cite{gal2022image} metrics. }
\label{tab:1}
\small 
\setlength{\tabcolsep}{8pt} 
\begin{tabular}{lcccc}
\toprule
\textbf{Metrics\textbackslash{}Methods} & DDS~\cite{hertz2023delta} & GSEditor~\cite{chen2023gaussianeditor} & \textsc{VcEdit} \\
\midrule
User Study                        & 1.57\% & 34.49\%           & \textbf{63.93\%} \\
CLIP similarity\cite{gal2022stylegan}                           &0.1470                &0.1917                    &\textbf{0.2108} \\
\bottomrule
\end{tabular}
\end{table}

\subsection{Quantitative Comparisons}
\label{sec:quanti}

Despite the subjective nature of 3D scene generation and editing, we adhere to established practices by utilizing the CLIP~\cite{radford2021learning} text-to-image directional similarity metric~\cite{gal2022stylegan} for quantitative analysis. This metric quantifies the semantic differences between the original and edited 3DGS scenes in relation to their corresponding prompts. 
Furthermore, we carried out a user study that collected data from 25 participants in 11 samples, each participant tasked with selecting the most superior edit from the edited outcomes of all comparison methods considering alignment to the target prompt and visual quality.
Table~\ref{tab:1} presents the quantitative results of the evaluation of our approach comparing with DDS and GSEditor in various test scenes. \textsc{VcEdit} not only achieves superior results in the user study but also outperforms in CLIP text-to-image directional similarity. This performance indicates that \textsc{VcEdit} provides high-quality edits that are closely aligned with the intended prompts, significantly outpacing the baseline methods in terms of editing quality and effectiveness.

\subsection{Editing with Diverse Source} 

Fig.~\ref{fig:extensive} highlights our \textsc{VcEdit}'s adaptability in handling a wide range of challenging scenes and prompts, spanning from intricate face and object transformations to extensive adjustments in large-scale scenes.
In the first row, our \textsc{VcEdit} produces an editing result that vividly presents the facial details of the target, ``grandma''. 
The second example highlights \textsc{VcEdit}'s ability to transform a small ``doll'' into a ``robot'' while carefully preserving the doll's key characteristics. 
In the last row, \textsc{VcEdit} demonstrates its capacity for large-scale scene modifications, convincingly altering a woodland scene to depict it as ``on fire''. These instances affirm \textsc{VcEdit}'s competency in delivering high-quality edits across a diverse array of tasks.

\section{Ablation and Discussion}
\label{sec:ablation}

\subsection{Ablation on Consistency Modules}
\label{sec:ablation_consist}
Sec.~\ref{sec:acc} and~\ref{sec:pcc} respectively introduced the Cross-attention Consistency Module (CCM) and the Editing Consistency Module (ECM). 
This section presents an ablation study to assess the impact of these modules. Given the critical role of multi-view consistent guidance in image-guided 3D editing as discussed in Sec.~\ref{sec:intro}, our analysis focuses on the view-consistency of the edited \textbf{2D} images. To understand why \textsc{VcEdit} outperforms the baseline methods, we revisit the cases of ``clown'' and ``brown bear'' from Sec.~\ref{sec:quali} for a detailed analysis.

\begin{figure}[t]
\centering
\includegraphics[width=0.99\linewidth]{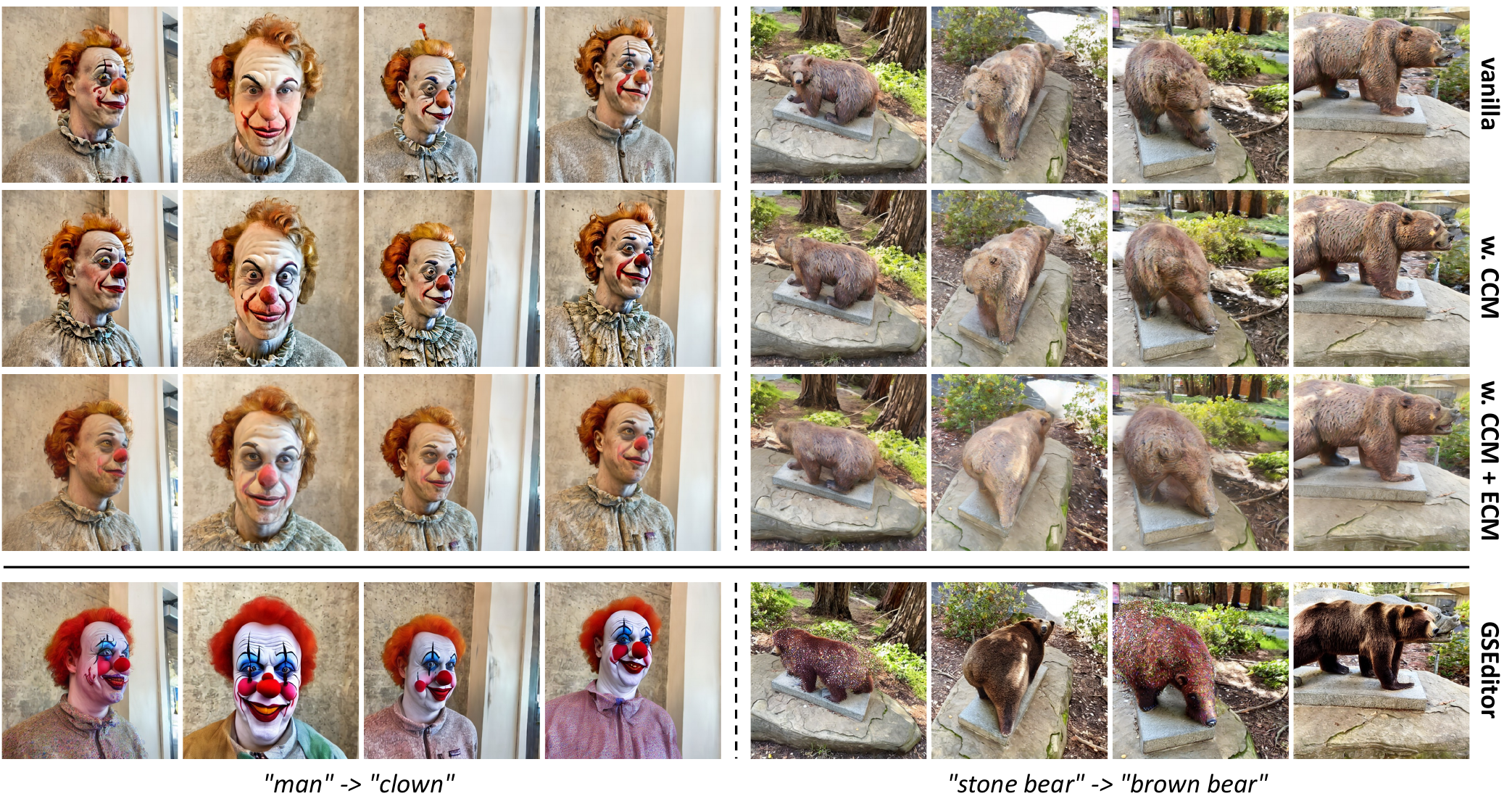}
\caption{\textit{Top}: 2D edited guidance images using three variant versions of \textsc{VcEdit}: vanilla version without any consistency module, version without ECM, and the original \textsc{VcEdit} employs both CCM and ECM.
\textit{Bottom}: 2D edited guidance images using GSEditor}
\label{fig:ablation}
\end{figure}

We conducted a comparative analysis of the edited images produced by three variant versions of \textsc{VcEdit}: the vanilla version \textit{without any consistency module}; the version \textit{with CCM but without ECM}; and the version that \textit{employs both CCM and ECM}.
The \textit{top 3 rows} in Fig.~\ref{fig:ablation} displays the comparison results:
(1) The edited images of the vanilla version contain significant view inconsistencies in the two samples. 
These incoherent edited contents among views can lead to the anomalies in the 3DGS.
(2) Applying the CCM markedly enhances the view consistency. Compared to the vanilla version, the edited images show a reduction of the incoherent content.
(3) When both consistency modules are employed, the results are in perfect coherence among the edited views.
In conclusion, both the Cross-attention and Editing Consistency Modules play vital roles in addressing multi-view inconsistency, enabling \textsc{VcEdit} to provide more accurate and detailed 3D editing through consistent image guidance.

In addition, we further compare the edited images produced by our \textsc{VcEdit} and GSEditor in the \textit{bottom 2 rows} of Fig.~\ref{fig:ablation}. The edited guidance images of our \textsc{VcEdit} employing both CCM and ECM achieves significantly better multi-view consistency than that of using GSEditor, which explains the performance gap in edited 3DGS discussed in Sec.~\ref{sec:quali}.

\noindent
\subsection{Ablation on Iterative Pattern}
\begin{figure}[t]
\centering
\includegraphics[width=0.99\linewidth]{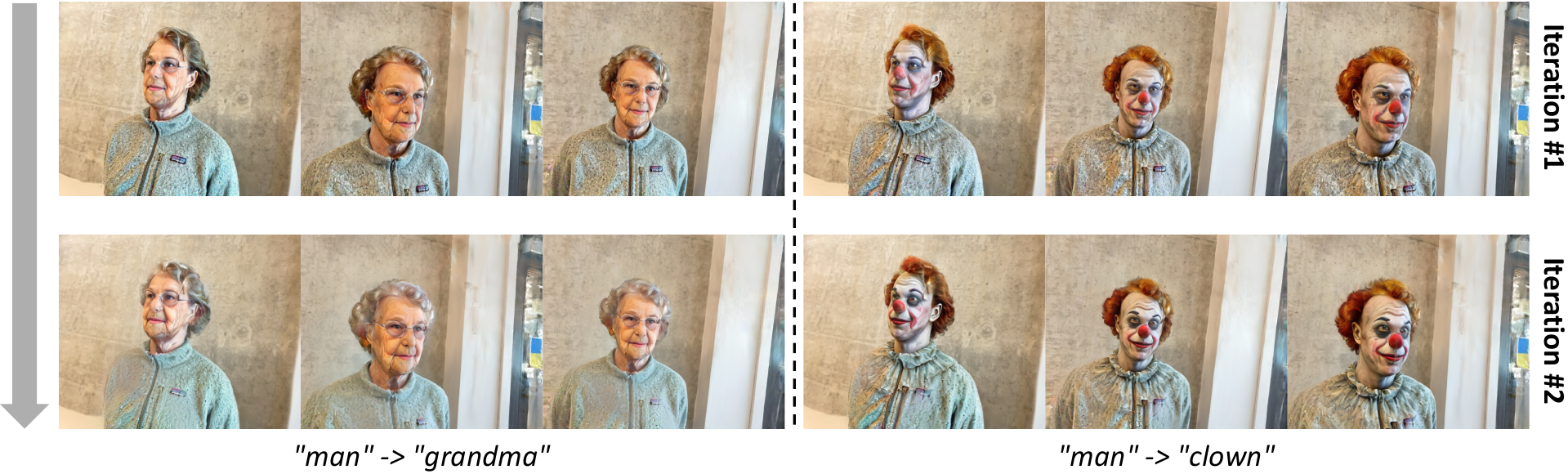}
\caption{Edited 3DGS after each iteration: Following the iterative pattern, the edited 3DGS is progressively refined by each iteration.}
\label{fig:ablation_iter}
\end{figure}

In Sec.~\ref{sec:iter}, we expand our framework to include an iterative pattern to further reduce the adverse effects of multi-view inconsistency. This section analyzes the impact of this iterative approach on the 3DGS editing performance. We specifically re-examine the ``man to grandma" and ``man to clown" examples by comparing the 3DGS editing outcomes of the first and second iterations. Our findings, presented in Fig.~\ref{fig:ablation_iter}, show that:
(1) In the ``man to grandma" case, the second iteration significantly refines facial details and clothing.
(2) In the ``man to clown" case, the second iteration further accentuates the ``clown" features by intensifying the makeup and smoothing facial details.
(3) Even in the first iteration, our \textsc{VcEdit} achieves relatively satisfying multi-view consistency, which is attributed to our effective Consistency Modules.
These enhancements confirm that through reciprocal and continuous refinement of 3DGS and image guidance, the iterative pattern progressively steers the samples towards a consistent style that aligns with the desired target, leading to improved editing quality.

\subsection{Limitations}
In line with earlier efforts in 3D editing~\cite{haque2023instructnerf2nerf,chen2023gaussianeditor} that utilize the image-guided editing pipeline, our \textsc{VcEdit} depends on the 2D editing models to generate image guidance, which leads to two specific limitations:
(1) Current diffusion-based image editing models occasionally fail to deliver high-quality image editing for intricate prompts, thus impacting the efficacy of 3D editing.
(2) In non-rigid editing scenarios that request drastically changing of object shape, current diffusion models produce highly variant editing results among views, which makes it hard for our consistency modules to rectify the tremendous inconsistency, thereby limits the quality of 3D editing.

\section{Conclusion}

In this paper, we propose the View-consistent Editing (\textsc{VcEdit}) framework for 3D Editing using text instruction, aiming to address the multi-view inconsistency issue in image-guided 3DGS editing.
By seamlessly incorporating 3DGS into the editing processes of the rendered images, \textsc{VcEdit} ensures the multi-view consistency in image guidance, effectively mitigating mode collapse issues and achieving high-quality 3DGS editing.
Several techniques are proposed to achieve multi-view consistency in the guidance images, including
two innovative Consistency Modules integrated in an iterative pattern.
Our extensive evaluations across various real-world scenes have demonstrated \textsc{VcEdit}'s ability to outperform existing state-of-the-art methods, thereby setting a new standard for text-driven 3D model editing. The introduction of \textsc{VcEdit} solves a critical problem in 3DGS editing, inspiring the exploration of future efforts.

\noindent
\textbf{Acknowledgements.} This research is supported by the National Research Foundation, Singapore under its AI Singapore Programme (AISG Award No: AISG2-GC-2022-005). 
Xuanyu Yi is supported by the Agency for Science, Technology AND Research. Long Chen was supported by HKUST Special Support for Young Faculty (F0927) and HKUST Sports Science and Technology Research Grant (SSTRG24EG04).
We also appreciate Ms. Sijing Lin and Dr. Zhongqi Yue for their contributions to the paper writing.


%
%
\bibliographystyle{splncs04}
\bibliography{main}

\clearpage

\section*{Appendix: Overview}

In the supplementary material, we provide more details of our algorithm and the implementation in Sec.~\ref{supp_sec:imple}. 
Additionally, further discussion regarding the two Consistency Modules of \textsc{VcEdit} is available in Sec.~\ref{supp_sec:discuss}. 
An extensive qualitative evaluation, including comparisons with NeRF-based editing methods, is conducted in Sec.~\ref{supp_sec:more_comp}.
The settings of the user study are detailed in Sec.~\ref{supp_sec:userstudy}.
Lastly, more visual results, extending the qualitative analysis of \textsc{VcEdit}, are presented in Sec.~\ref{supp_sec:more_res}.

\begin{algorithm}[ht]
\scriptsize
	\caption{Detailed Pipeline of \textsc{VcEdit} in \textbf{One} Iteration}
	\textbf{Input:} 
 
    \hspace{0.9cm}
    \begin{tabular}{lcr}
    U-Net of Conditional Diffusion Model $\eps_\theta(\cdot)$&&\\
    Sequence of timesteps $T = t_1 > t_2 > \ldots > t_N = 1$ &&\\
    Source 3DGS $\gsrc$ &&\\
    Source/target prompts as conditions $y^\text{src} (\text{e.g., ``man''}), y^\text{tgt}$ (\text{e.g., ``clown''})&&\\
    \end{tabular}
 
	\begin{algorithmic}[1]

	\State \textbf{3DGS $\to$ Images} (Input: $\gsrc$ $\to$ Output: $\Isrc$)

    \State $\;\;\;$ Rendering 3DGS to images: $\Isrc = \mathcal{R}(\gsrc)$
    
	$\;$
    \State \textbf{Multi-view Images Editing} (Input: $\Isrc$ $\to$ Output: $\Iedt$)

        \State $\;\;\;$ Encoding images to latents: $\zs, \zo = \mathcal{E}(\Isrc)$
	
		\State $\;\;\;$ \textbf{for} $t=t_1,t_2,\ldots t_N $ \textbf{do}
		

        \State $\;\;\;\;\;\;$ Sample noise: $\eps \sim \mathcal{N}(0, \mathcal{I})$

        \State $\;\;\;\;\;\;$ Add noise: $z_t = \sqrt{\alpha_t}\zs + \sqrt{1 - \alpha_t} \eps$

        \State $\;\;\;\;\;\;$ Add noise to original latents: $\zo_t = \sqrt{\alpha_t}\zo + \sqrt{1 - \alpha_t} \eps$


        \State $\;\;\;\;\;\;$ Noise prediction by \textcolor{blue}{\textbf{CCM}}-applied U-Net: $\eps_\theta, \eps^{\text{ori}}_\theta = \eps_\theta(z_t, y^\text{tgt}), \eps_\theta(\zo_t, y^\text{src})$

        \State $\;\;\;\;\;\;$ Compute a noise offset proposed in~\cite{xu2023inversionfree}: $\Delta\eps = (\zo_t - \sqrt{\alpha_t}\zo) / \sqrt{1 - \alpha_t}$

        \State $\;\;\;\;\;\;$ Obtain the edited latents: $\ze = (\zs - \sqrt{1 - \alpha_t}(\eps_\theta - \eps^{\text{ori}}_\theta + \Delta\eps)) / \sqrt{\alpha_t}$


        \State \textcolor{blue}{$\;\;\;\;\;\;$ Copy a 3DGS from source (\textbf{ECM} starts): $\gft = \text{copy($\gsrc$)}$}

        \State \textcolor{blue}{$\;\;\;\;\;\;$ Fine-tune the 3DGS: $\gft = \: \underset{\G}{\mathrm{argmin}} \: \sum_{v \in \Vs}{ \mathcal{L}_{\text{edit}}(\mathcal{R}(\G, v),\: \dec(\ze_{v}))}$}

        \State \textcolor{blue}{$\;\;\;\;\;\;$ Render and encode to latents (\textbf{ECM} ends): $\zc = \mathcal{E}(\mathcal{R}(\gft, v)), \; v \in \Vs$}


        \State \textcolor{blue}{$\;\;\;\;\;\;$ Blend by the \textbf{CCM}'s consistent mask: $\zb = \mc * \zc  + (1 - \mc) * \zs$}

        \State $\;\;\;\;\;\;$ Next timestep: $\zs = \textcolor{blue}{\zb}$
		
		\State $\;\;\;$ \textbf{end for}

        \State $\;\;\;$ Decoding latents to edited guidance images: $\Iedt = \mathcal{D}(\zb)$
	   
	 $\;$ 

\State \textbf{Images $\to$ 3DGS} (Input: $\Iedt$ $\to$ Output: $\gedt$ )

\State $\;\;\;$ Fine-tune the source 3DGS: $\gedt = \: \underset{\G}{\mathrm{argmin}} \: \sum_{v \in \Vs}{ \mathcal{L}(\mathcal{R}(\G, v),\: \Iedt)}$
    
	\end{algorithmic}
	\textbf{Output: } The final edited 3DGS $\gedt$.
\label{algo:1}
\end{algorithm}

\section{More Details of our \textsc{VcEdit}}
\label{supp_sec:imple}

In the \textit{Our Method} section of the main paper, we provide an overview of \textsc{VcEdit} along with the introductions of our two Consistency Modules. In this section, we provide more details in the algorithm and implementation of \textsc{VcEdit} with a step-by-step demonstration in Algorithm~\ref{algo:1}, where our two consistency modules are highlighted in \textcolor{blue}{BLUE} color.

\subsection{More Details in our Overall Pipeline}
As outlined in the \textit{Preliminary} and \textit{Our Method} sections, \textsc{VcEdit} employs an iterative image-guided 3DGS editing pipeline that takes user-specified text prompt as instruction.
Initially, the source 3DGS ($\gsrc$) are rendered to images ($\Isrc$) from various views (\textit{line 1--2} in Algorithm~\ref{algo:1}). 
Subsequently, employing our specially designed multi-view image editing process based on InfEdit~\cite{xu2023inversionfree} (\textit{line 3--18} in Algorithm~\ref{algo:1}), a set of multi-view consistent edited images ($\Iedt$) are generated and employed as image guidance to fine-tune $\gsrc$ (\textit{line 19--20} in Algorithm~\ref{algo:1}). 

During fine-tuning, we adhere to the methodology proposed by the GSEditor~\cite{chen2023gaussianeditor}, incorporating both a MAE loss and a VGG-based LPIPS loss~\cite{simonyan2014very, chen2023gaussianeditor}.
Additionally, we employ the HGS regularization, as suggested by GSEditor~\cite{chen2023gaussianeditor}, which limits the positional shifts of the Gaussian induced by densification to maintain the essential information in $\gsrc$. Consequently, the training objective is defined as:
\begin{equation}
    \mathcal{L} = \lambda\mathcal{L}_{\text{MAE}}(\mathcal{R}(\G, v), \Iedt) + \lambda\mathcal{L}_{\text{LPIPS}}(\mathcal{R}(\G, v), \Iedt)
    + \sum\lambda_j(P_j - \hat{P}_j)
\end{equation}
Here, $P_j$ indicates the original position of the $j$-th Gaussian in $\mathcal{G}$, while $\hat{P}_j$ denotes its shifted position during training.
In \textsc{VcEdit}, we set the $\lambda$ to 10 and each $\lambda_i$ to 50. 
After 400 steps of fine-tuning, $\gsrc$ transforms into the edited version, $\gedt$. 
In subsequent iterations, $\gedt$ becomes the new $\gsrc$, and the editing cycle continues.

\subsection{More Details in our Multi-view Images Editing Process}
In the main paper, we simplified the explanation of our image editing procedure to enhance reader comprehension.
This section elaborates on the process, integrating it with our baseline image editing framework,  InfEdit~\cite{xu2023inversionfree} (\textit{line 3--18} in Algorithm~\ref{algo:1}).

Initially, in \textit{Our Method} section, we introduce our image editing process as a multi-timestep cycle, where each timestep's process is represented by $\zs \rightarrow \ze \rightarrow \zc \rightarrow \zb$.
Following InfEdit~\cite{xu2023inversionfree}, our detailed implementation introduces an additional set of \textbf{original latents}, $\zo$, extracted from the original images (\textit{line 4} in Algorithm~\ref{algo:1}).
These latents are utilized by InfEdit in the add-noise and denoising process (\textit{line 8--11} in Algorithm~\ref{algo:1}) and does not participate in any of our Consistency Module.

Moreover, InfEdit's U-Net architecture~\cite{xu2023inversionfree} is partitioned into three branches: the \textit{Source Branch}, the \textit{Layout Branch}, and the \textit{Target Branch}. These branches are interconnected by two sets of cross-attention maps, which are denoted as $\m = \{\m^\text{source-layout}, \m^\text{target}\}$. Within our Cross-attention Consistency Module (CCM), we execute inverse rendering and subsequent re-rendering for both cross-attention map sets, ensuring uniform editing outcomes across all branches.

Our Editing Consistency Module is activated for every 5 timesteps for a more efficient forwarding. Since the $\ze$ in low-resolution does not satisfy the requirement of 3DGS fine-tuning, we initially decode them to images and use the images to fine-tune the 3DGS $\gft$ (\textit{line 13} in Algorithm~\ref{algo:1}). After a rapid fine-tuning, we render the $\gft$ back to images, and encode the rendered images to obtain $\zc$ (\textit{line 14} in Algorithm~\ref{algo:1}).

\section{Further Discussion on Consistency Modules}
\label{supp_sec:discuss}

\begin{figure}[t]
\centering
\includegraphics[width=0.99\linewidth]{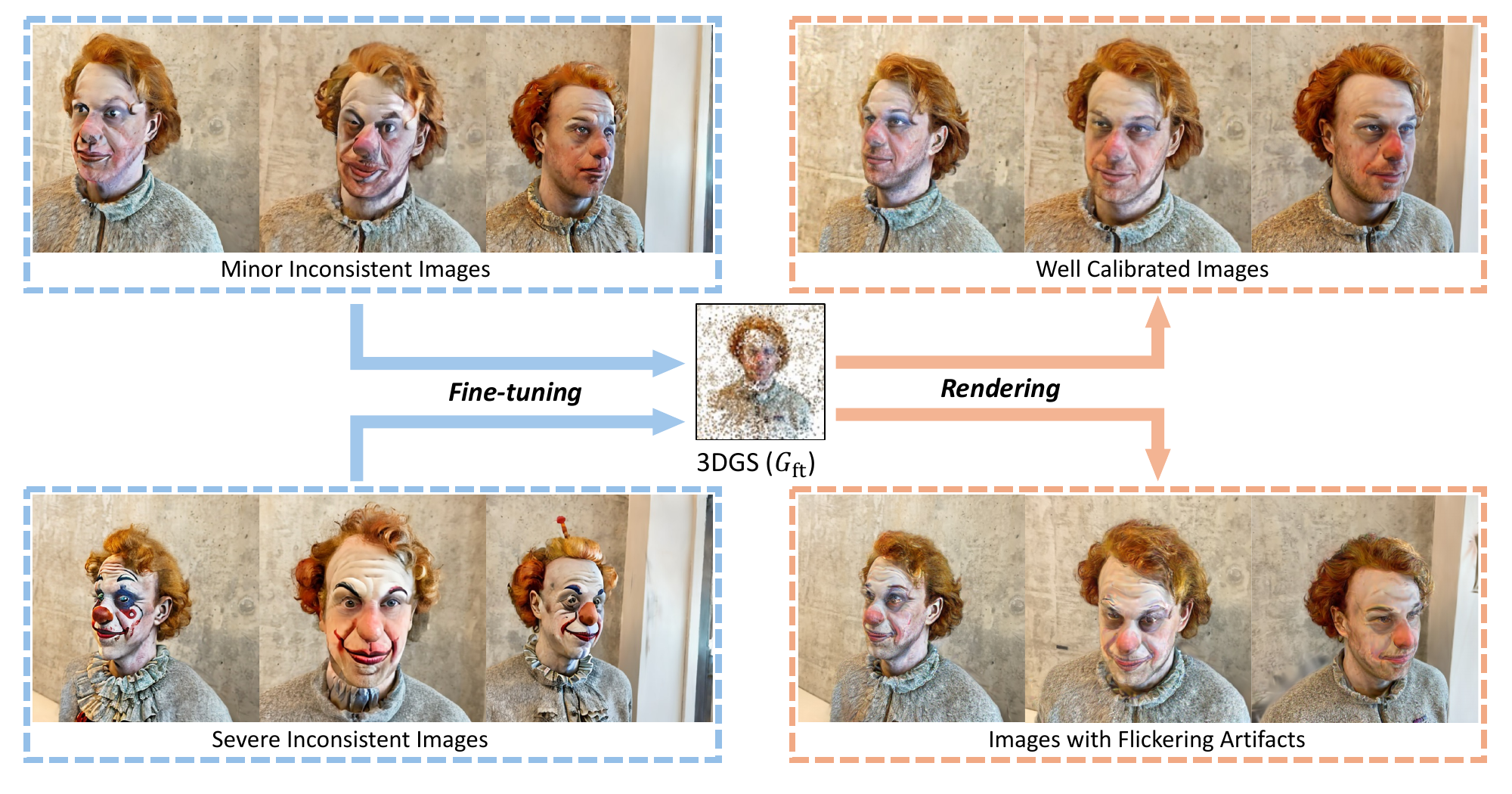}
\caption{Calibration Capability of 3DGS: A set of multi-view inconsistent images (\textit{left}) serves as guidance for fine-tuning a 3DGS model, which is then re-rendered into images (\textit{right}). 
The 3DGS model has the capability to calibrate minor inconsistency (\textit{top}) but will exhibit flickering artifacts when trained with severe inconsistent images (\textit{bottom}).
}
\label{fig:calibration}
\end{figure}

\subsection{Discussion on Editing Consistency Module}

In \textsc{VcEdit}, we introduce an innovative Editing Consistency Module (ECM) that leverages the subtle calibration potential of the 3DGS model to correct minor inconsistencies arising at each timestep of the image editing process. This section presents comparative experiments designed to further investigate the calibration capabilities and elucidate the effectiveness of our ECM.

In Fig.~\ref{fig:calibration}, we employ two sets of images, each exhibiting different levels of multi-view inconsistency, to guide the fine-tuning of a 3DGS. This model is then rendered back into images (\textit{right}). The \textit{top} row illustrates the model's capability to correct minor multi-view inconsistencies effectively, as shown by the absence of mode collapse in the rendered images (\textit{top-right}). In contrast, the \textit{bottom} row reveals the model's limitations when confronted with severe inconsistencies, which lead to poorly calibrated images marked by flickering artifacts (\textit{bottom-right}). 

This observation elucidates the efficacy of our Editing Consistency Module in addressing inconsistencies by continually calibrating minor discrepancies at each image editing timestep, while directly employing original multi-view inconsistent images as guidance results in mode collapse. 

\subsection{Discussion on Cross-attention Consistency Module}
In the main paper, we introduce the Cross-attention Consistency Module (CCM) as a novel approach to synchronize the attentive regions across all views within the U-Net layers, thereby facilitating the U-Net to produce multi-view consistent predictions.
Specifically, the cross-attention maps from all views are consolidated through a process of inverse-rendering (2D $\to$ 3D) and subsequent rendering (3D $\to$ 2D).

\begin{figure}[t]
\centering
\includegraphics[width=0.99\linewidth]{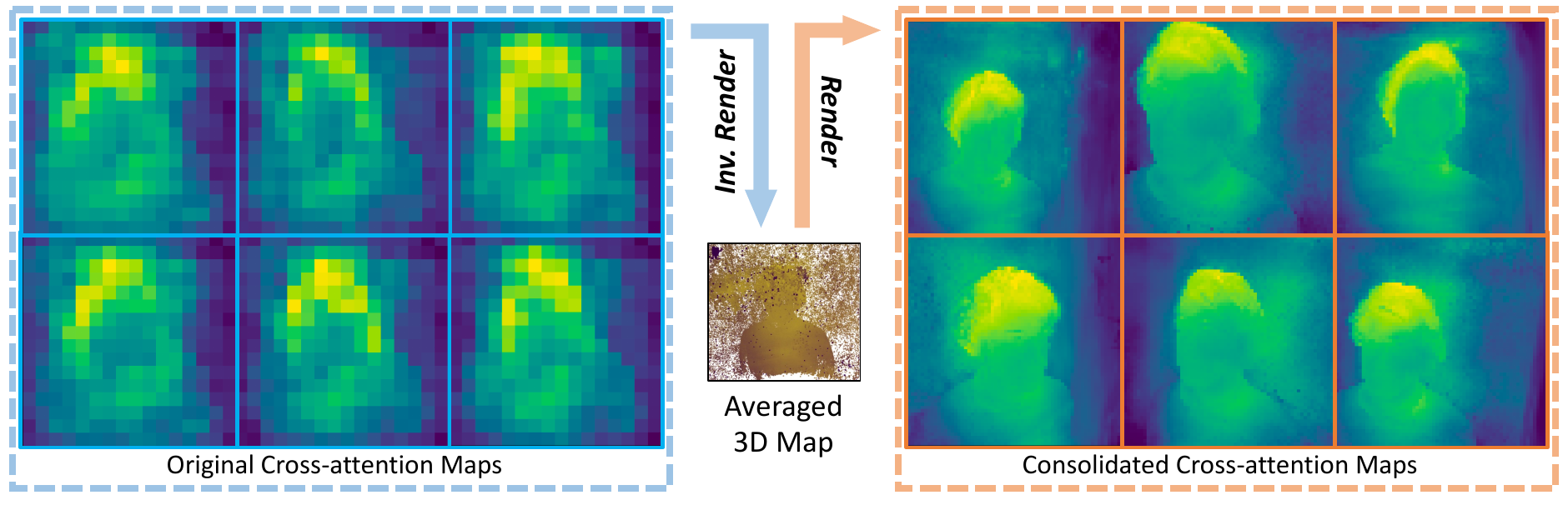}
\caption{Consolidation Effect of Our CCM on the ``man $\to$ clown'' sample: Initially, incoherent multi-view cross-attention maps (\textit{left}) undergo an inverse rendering process into 3D space, resulting in an averaged 3D map. This 3D map is subsequently rendered back into the respective views, yielding consolidated cross-attention maps (\textit{right}), which provides unified attention weights across various views to each 3D region.
}
\label{fig:crossattn}
\end{figure}

Fig.~\ref{fig:crossattn} showcases the impact of our CCM on the ``man $\to$ clown'' example, comparing multi-view cross-attention maps towards the term ``clown'' before and after the consolidation process. 
The comparison reveals that, unlike the original cross-attention maps on the \textit{left}, which are coarse and inconsistent across views, the consolidated maps exhibit a refined, view-consistent attention on each 3D region.

\begin{figure}[t]
\centering
\includegraphics[width=0.99\linewidth]{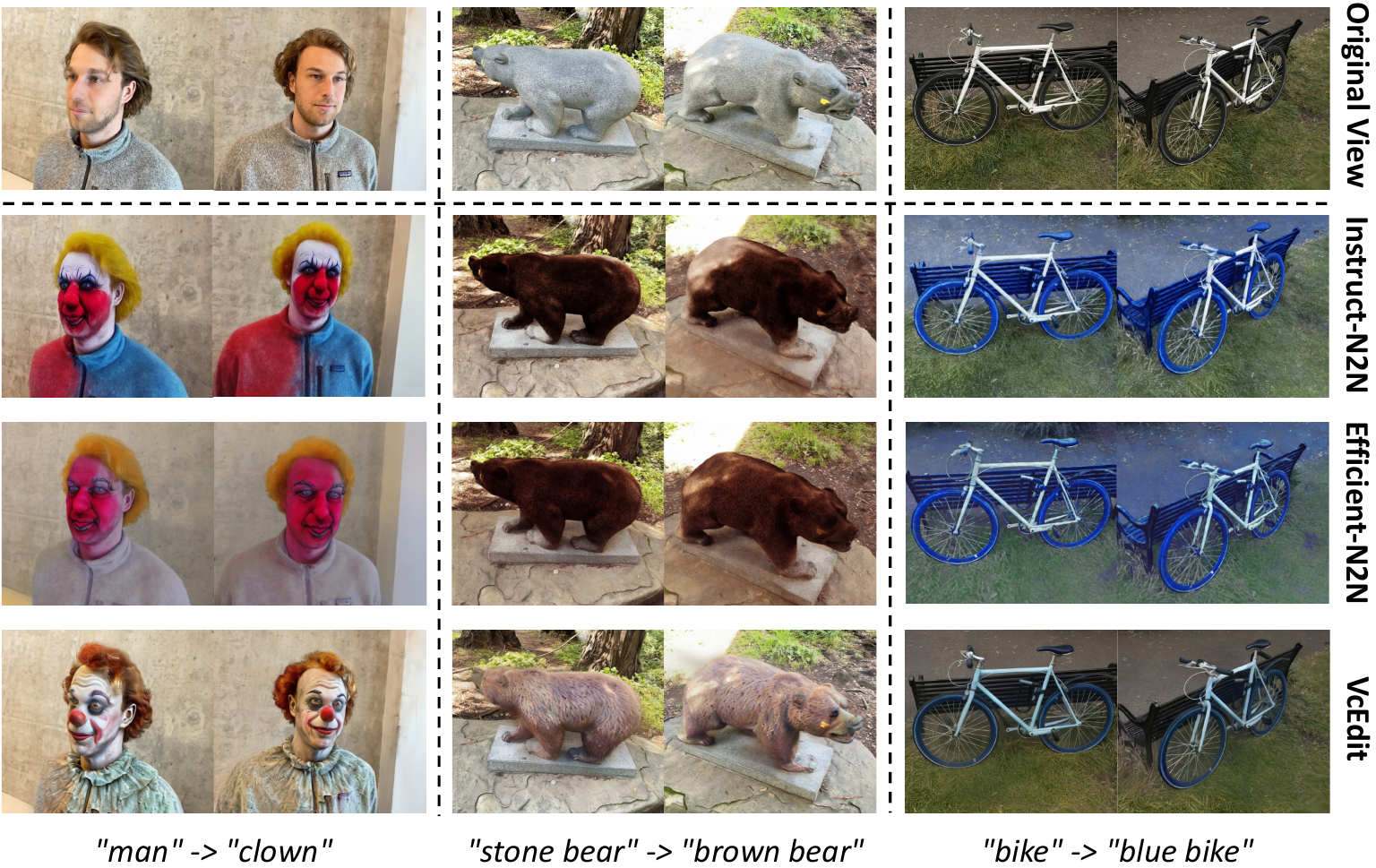}
\caption{Qualitative Comparison with the Instruct-NeRF2NeRF~\cite{haque2023instructnerf2nerf} and the Efficient-NeRF2NeRF~\cite{song2023efficient}: Our \textsc{VcEdit} outperforms both the NeRF editing methods in terms of achieving finer editing details. Conversely, the NeRF editing approaches produce overly smooth outcomes along with unintended colorization in the background.
}
\label{fig:nerf}
\end{figure}

\section{Qualitative Comparison with NeRF Editing Methods}
\label{supp_sec:more_comp}

In our main paper, we compare the editing quality of \textsc{VcEdit} with prevalent 3DGS editing methods, demonstrating that \textsc{VcEdit} significantly surpasses existing state-of-the-art methods in terms of editing quality.
To expand our analysis, this section introduces a comparison with NeRF editing approaches, with the findings depicted in Fig.~\ref{fig:nerf}.
Specifically, we contrast \textsc{VcEdit} with the widely recognized Instruct-NeRF2NeRF\cite{haque2023instructnerf2nerf} and the innovative Efficient-NeRF2NeRF~\cite{song2023efficient}.


The comparative analysis depicted in Fig.~\ref{fig:nerf} reveals that NeRF editing methods tend to yield overly smoothed outcomes, in contrast to \textsc{VcEdit}, which delivers results rich in detail.
Furthermore, the outcomes of both NeRF editing methods exhibit unintended colorization across the scene, indicating a failure to maintain the original background's integrity. This limitation stems from a fundamental challenge inherent to NeRF models: Compared with 3DGS model, it is more difficult for NeRF to achieve precise local editing.
Collectively, these observations underscore \textsc{VcEdit}'s superior performance in 3D editing tasks.
\section{Detailed Settings of User Study}
\label{supp_sec:userstudy}

\begin{figure}[h]
\centering
\includegraphics[width=0.82\linewidth]{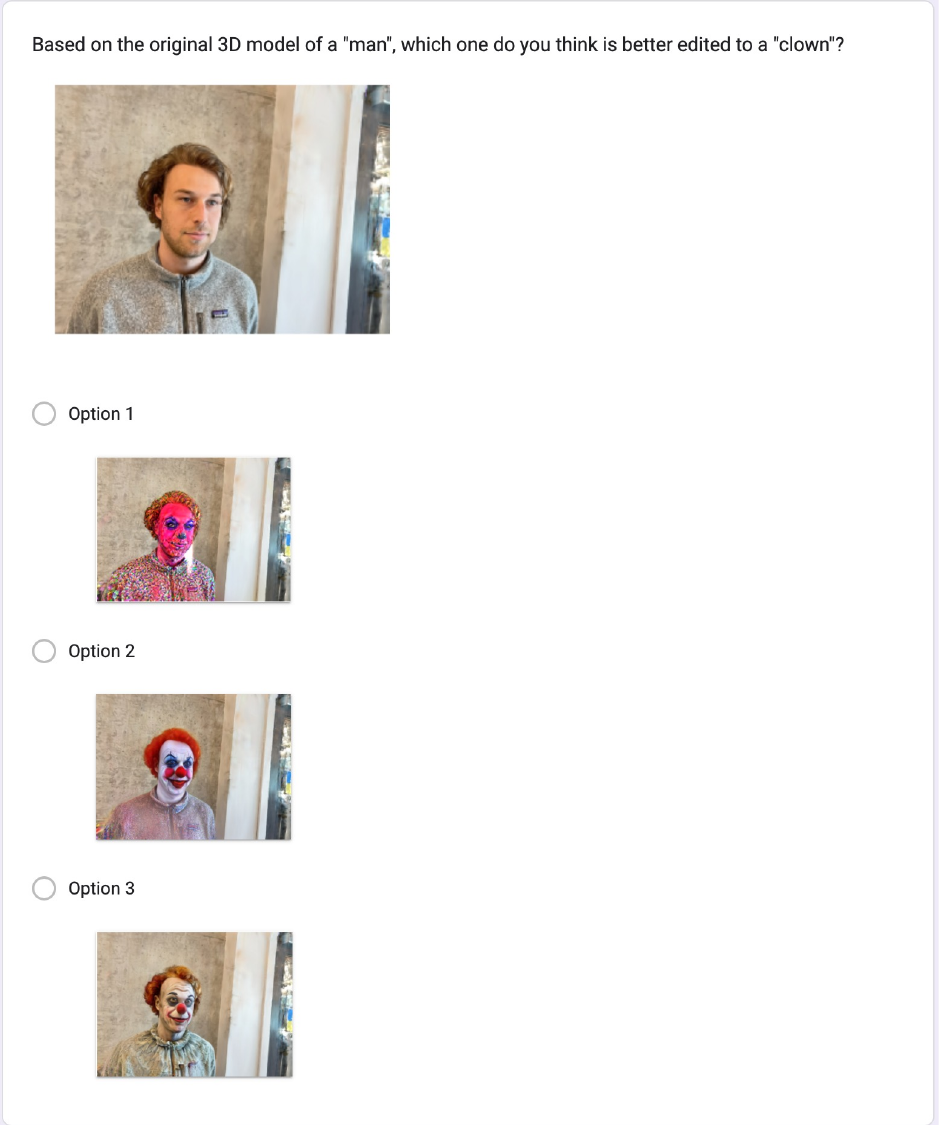}
\caption{
 Question in our User Study: Participants were asked to select one editing they deemed in the best quality.
}
\label{fig:user_study}
\end{figure}

In our user study, participants encountered a series of questions, each comprising one original view and three corresponding rendered views from the 3DGS edited via different comparative methods. An illustrative example of such a question is presented in Fig.~\ref{fig:user_study}, where participants were tasked with selecting the editing they considered to be of the highest quality. To promote impartiality in responses, the sequence of the methods was randomized for each question, and all options were presented anonymously.

\begin{figure}[h]
\centering
\includegraphics[width=0.99\linewidth]{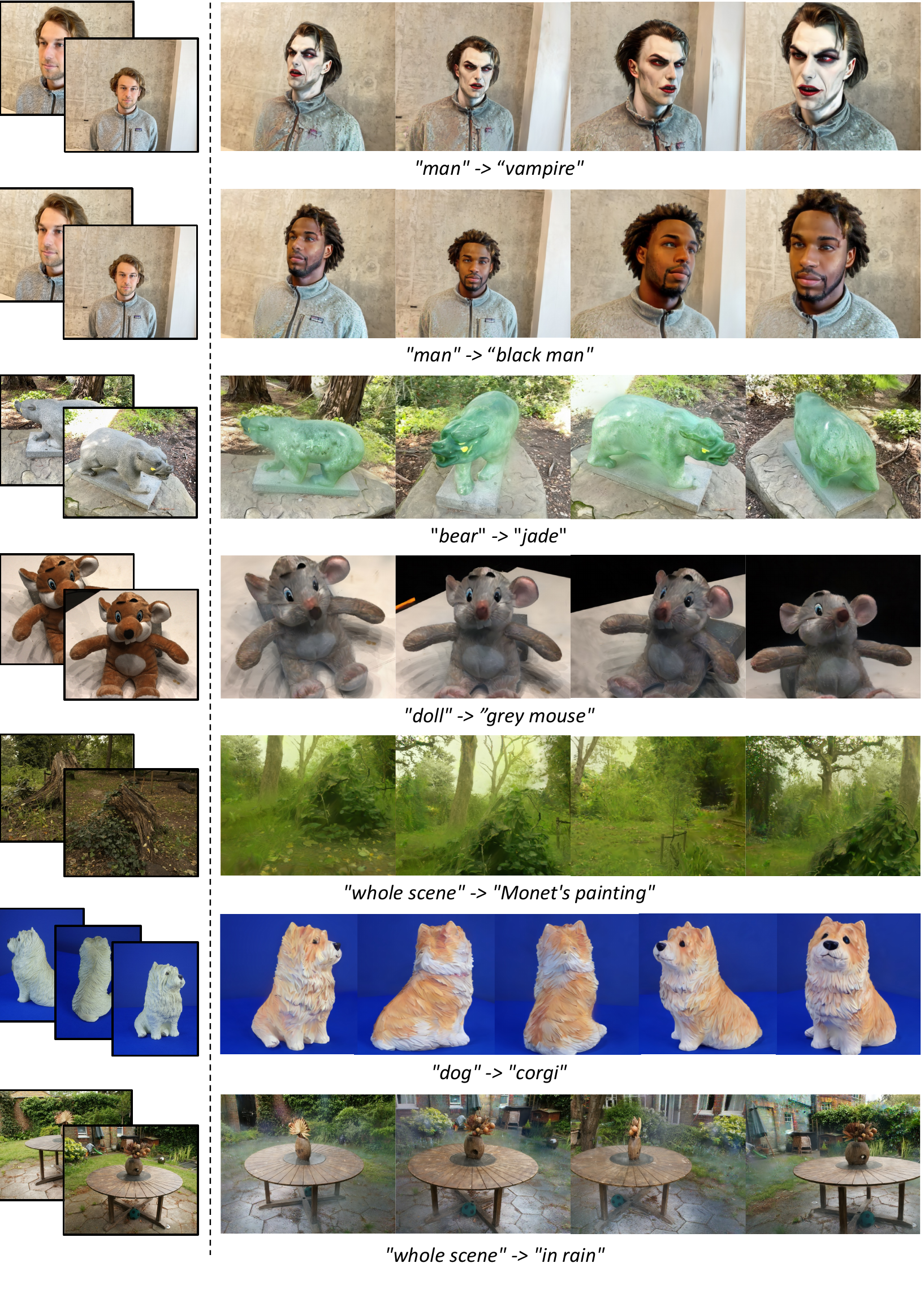}
\caption{
 More Editing Results of \textsc{VcEdit}: Our method is capable of various editing tasks, including face, object, and large-scale scene editing. The \textit{leftmost column} demonstrates the original view, while the \textit{right four columns} show the rendered view of edited 3DGS.
}
\label{fig:more_res}
\end{figure}

\section{More Editing Results of \textsc{VcEdit}}
\label{supp_sec:more_res}

As the supplementary to our main paper, we present more editing result produced by \textsc{VcEdit} in Fig.~\ref{fig:more_res}.
Further editing outcomes generated by \textsc{VcEdit} are displayed in Fig.~\ref{fig:more_res}, where our \textsc{VcEdit} produces high-quality editing result for each sample.
These results underscore the adaptability of \textsc{VcEdit} in managing a diverse array of complex scenarios and prompts, ranging from detailed facial and object modifications to broad-scale scene alterations.

\end{document}